\documentstyle[12pt]{article}
\begin{document}
\pagestyle{plain}
\setcounter{page}{1}
\baselineskip16pt

\def\kth{$k^{\rm th}$ }
\def\Tr{{\rm Tr}\,}
\def\pvint{-\!\!\!\!\!\!\int}
\def\equno#1{Eq.~(\ref{#1})}
\def\sectno#1{section~\ref{#1}}
\def\figno#1{Fig.~(\ref{#1})}
\def\qg{quantum gravity}
\def\bu{baby universe}

\begin{titlepage}

\begin{flushright}
PUPT-1422\\
hep-th/9310098
\end{flushright}
\vspace{20 mm}

\begin{center}
{\huge Scaling Functions for Baby Universes}

\vspace{5mm}

{\huge in Two-Dimensional Quantum Gravity}
\end{center}

\vspace{10 mm}

\begin{center}
{\large Steven S. Gubser and Igor R. Klebanov}

\vspace{3mm}

Joseph Henry Laboratories\\
Princeton University\\
Princeton, New Jersey 08544

\end{center}

\vspace{2cm}

\begin{center}
{\large Abstract}
\end{center}

We apply the recently proposed transfer matrix formalism to 2-dimensional
\qg\ coupled to $(2, 2k-1)$ minimal models. We find that the propagation
of a parent universe in geodesic (Euclidean) time is accompanied by
continual emission of \bu s and derive a distribution function
describing their sizes. The $k\to \infty~ (c\to -\infty)$
limit is generally thought to correspond to classical geometry, and we
indeed find a classical peak in the universe distribution function. However,
we also observe dramatic quantum effects associated with \bu s at
finite length scales.

\vspace{2cm}
\begin{flushleft}
October 1993
\end{flushleft}
\end{titlepage}
\newpage

\section{Introduction}
\label{Intro}

One fascinating aspect
of \qg\ is the possibility of topology changing processes where
a compact connected 3-geometry (a ``baby universe'') splits off or joins
a large ``parent universe.'' Such processes, peculiar as they may seem,
are quite natural at the Planckian distance scales, where geometry
undergoes large quantum fluctuations. Even if the topology changing
processes are confined to the smallest distance scales, they may
drastically affect the observed physics. As shown in ref.~\cite{CGS},
summation over all possible emissions and absorptions of the Planck-size baby
universes effectively gives rise to averaging over the fundamental constants
of nature. If the weight in the average is sharply peaked, then
the fundamental constants are determined \cite{SC}.
The wormhole theory of fundamental constants shows that,
via some unexpected connections, the short-distance properties of \qg\
can affect the universe at large in crucial ways. The status of this
theory when applied to 4-dimensional \qg\ is somewhat uncertain due to
the instability of the Euclidean path integral. It may be, for instance,
that the weight in the average over the coupling constants is not
sharply peaked \cite{US}. If, on the other hand, the weight is sharply
peaked, then it is not quite clear what prevents the emission and
absorption of macroscopic baby universes \cite{VK}. While these puzzles await
their resolution through a better understanding of 4-dimensional
\qg, we may try to gain some intuition about the
topology changing processes from low-dimensional models.

The 2-dimensional Euclidean \qg\ coupled to $c\leq 1$ conformal matter appears
to provide some promising toy models which are not plagued by
any perturbative instabilities. Our understanding of these theories has
dramatically improved recently, in part due to the success of the matrix
model techniques \cite{GM}. With their help we can perform summations over
discretized random surfaces and extract the exact results in the
universal continuum limit. Until recently, however, the available exact
solutions have not shed any light on the physics of
topology changing processes. The problem is that the matrix models do
not give any direct information on the internal geometry of
2-dimensional space-times. Some first pieces of information were
extracted via direct Monte Carlo studies of triangulated random
surfaces \cite{AM}.
It was noted that the set of points at a given geodesic distance
$D$ from some point $P$ typically consists of many disconnected
loops, whose number grows rapidly with $D$. The study in ref.~\cite{AM}
gave a first indication that, as a 1-dimensional universe (a string)
propagates forward in geodesic (Euclidean) time, it is very likely
to continually emit baby universes. This phenomenon received a detailed
quantitative confirmation in a recent very interesting paper by
Kawai, Kawamoto, Mogami and Watabiki \cite{Kawai1}.
Relying on new combinatorial
techniques that they invented, these authors derived a formula for the
average number of loops with lengths between $L$ and $L+dL$ located at a
geodesic distance $D$ from some point on the surface,

\begin{equation}
\rho(L, D) dL =
{3\over 7 \sqrt{\pi} D^2} \left(x^{-5/2}+{1\over 2} x^{-3/2}+{14\over 3}
x^{1/2} \right) e^{-x} dL                                   \label{Scaling}
\end{equation}

\noindent
where $x=L/D^2$. A number of beautiful conclusions follow from this
formula. First of all, $\rho(L, D)$ is not normalizable at small $L$,
which means that the microscopic \bu s are overwhelmingly likely to be
emitted.
Secondly, apart
from an overall dimensionful factor, $\rho$ is a function of the
dimensionless scaling variable $x$. Introducing the moments,

\begin{equation}
\langle L^n \rangle_D=\int_0^\infty \rho(L, D) L^n dL \ ,
\end{equation}

\noindent
one finds that for $n<2$ they are dominated by the non-universal
short-distance cut-off.  But for $n\geq 2$ there are simple scaling
relations $\langle L^n \rangle_D \sim D^{2n}$,
whose form follows essentially from
$D$ having the dimension of (Length)$^{1/2}$.

The distribution function $\rho(L, D)$ is very important because it quantifies
the effects of \bu s in pure Euclidean 2-dimensional quantum gravity.
The next question is how the coupling to matter affects the topology
changing processes.  In this paper we begin to address this question by
considering the emission of baby universes
in 2-dimensional \qg\ coupled to the $(2, 2k-1)$ minimal
models, whose central charges are $c=1-3(2k-3)^2/(2k-1)$.
These theories have been identified \cite{Staud}
with the multicritical points of the one-matrix model \cite{Kaz},
\cite{GM}.
The $k=2$ theory corresponds to pure gravity, where
it suffices to consider the discretizations of random surfaces with
squares only. For $k=3~(c=-22/5)$ one needs squares and hexagons; for
$k=4$ one needs squares, hexagons and octagons; etc. For $k\geq 3$ some
polygons enter with negative weights, which is not surprising given that
these models are not unitary. We find that, for odd $k$, the negative
weights have so much effect that there is no sensible positive
$\rho(L, D)$. For even $k$, however, there does exist a positive
$\rho(L, D)$, whose calculation constitutes the main result of this
paper. We find that it depends on the scaling variable
$x=L/D^{1/(k-3/2)}$ and for small $x$ diverges as $x^{-k-1/2}$.
Thus, the emission of microscopic \bu s becomes more enhanced with
increasing $k$. For large $k$, $c\to -\infty$ and the sum over
surfaces is expected to be dominated by classical geometry. In this
limit we indeed find that $\rho(L, D)$ exhibits a sharp peak that
corresponds to the macroscopic classical geometry. Surprisingly, we
also find that baby universes remain prevalent for all length scales less
than a fixed constant, $s_c$, times the macroscopic (classical) length
scale, and calculate the critical value $s_c$.
As a further application of these ideas,
we discuss the production of \bu s by a very
large parent universe of length $L_0$. We find that their distribution
is governed by a very simple scaling law,

\begin{equation}
\rho (L_0\to \infty, L, D)\sim L_0 D L^{-k-1/2} \ ,
\end{equation}

\noindent
which applies to both even and odd $k$.

The structure of our paper is as follows.
In \sectno{MMloops} we review and rederive the necessary matrix
model results.  In \sectno{TransferM} we extend the transfer matrix
formalism of ref.~\cite{Kawai1} to general discretizations, which are
necessary to describe the theories with matter.  In \sectno{TubeAmp} we
calculate the transfer matrix in some limiting cases, and in
\sectno{BStructure} we study the  branching structure of space-time by
deriving the distribution functions for \bu s.  In \sectno{Discuss} we
conclude with a few remarks.

\section{The Disk Amplitudes}
\label{MMloops}

For the calculation of the transfer matrix, the only result needed from
the matrix models is the disk amplitude---that is, the partition
function for surfaces with one boundary loop.  It was
shown in \cite{MSS} that in the $k^{\rm th}$
multicritical theory, the universal part of the disk amplitude is

\begin{equation}
F_k(L, \tau) = {1 \over L} (\sqrt{\tau})^{k-1/2}
     K_{k-1/2} (\sqrt{\tau} L)                            \label{ExactFL}
\end{equation}

\noindent
where $\tau$ is the cosmological constant and $K_{k-1/2}$ is a modified
Bessel function.  We will need to keep track of certain non-singular parts
of the disk amplitudes because they will be important in the calculation
of the transfer matrix. Thus, we present our own calculation of
the disk amplitudes.

Throughout this paper, we will be
working with surfaces without handles, so
it is sufficient to use saddle point techniques in the
matrix models.
The partition function of the
\kth multicritical model (that is, the sum of the weights of all vacuum
diagrams, including the disconnected ones) is

\begin{equation}
{\cal Z} = \int d\Phi \exp \left[ -\beta \, \Tr V(\Phi ) \right]
\end{equation}

\noindent
where $\Phi$ is a $N \times N$ matrix with $N \sim \beta \to \infty$,
and

\begin{equation}
V(\Phi) = \sum_{m \geq 1} {c_m \over 2 m} \Phi^{2 m}
\end{equation}

\noindent
is a potential energy function.  (We restrict
ourselves to even $V(\Phi)$ because they are more easily analyzed,
but we believe that more general $V(\Phi)$ do not exhibit any more
general behavior.)

Let $u(\lambda)$ be the density of eigenvalues of $\Phi$, normalized so that
$\int d\lambda \, u(\lambda) = 1$.  As shown in \cite{IZ78}, $u(\lambda)$
has support $[-\sqrt{z},\sqrt{z}\,]$ for some $z$, and

\begin{equation}
V'(\lambda ) = 2 {N \over \beta} \pvint_{-\sqrt{z}}^{\sqrt{z}}
     d \mu {u(\mu) \over \lambda - \mu}\ .
\end{equation}

\noindent
For $\lambda \in {\bf C} - [-\sqrt{z},\sqrt{z}\,]$, define

\begin{equation}
F(\lambda ) = 2 {N \over \beta} \int_{-\sqrt{z}}^{\sqrt{z}}
     d \mu {u(\mu) \over \lambda - \mu} \ .
\end{equation}

\noindent
$F(\lambda)$ is analytic except for a branch cut along
$[-\sqrt{z},\sqrt{z}\,]$: for
$\lambda \in [-\sqrt{z},\sqrt{z}\,]$ and $\epsilon > 0$ infinitesimally
small, $F(\lambda \pm i \epsilon) =
V'(\lambda) \mp 2 \pi i (N/\beta) u(\lambda)$.  Also,
$F(\lambda) = 2 (N/\beta) / \lambda + O(\lambda^{-2})$ for large
$\lambda$.  It is not hard to prove that these properties uniquely
determine $F(\lambda)$ and hence $u(\lambda)$.  It turns out that
$F(\lambda)$ has the form

\begin{equation}
F(\lambda) = V'(\lambda) - f(\lambda) \lambda
    \sqrt{1 - {z \over \lambda^2}}
\end{equation}

\noindent
where $f(\lambda)$ is a polynomial.  Both $z$ and $f(\lambda)$ are
determined by the large $\lambda$ expansion of $F(\lambda)$.  Defining
for convenience the functions $g(\lambda) = \lambda V'(\lambda)$ and
$h(\lambda) = \lambda^2 f(\lambda)$, we find that

\begin{equation}
g(\lambda) \left(1 - {z \over \lambda^2} \right)^{-1/2} =
   \left( \sum_{m \geq 1} c_m \lambda^{2 m} \right)
   \left( \sum_{m \geq 0} {(2 m)! \over m!^2 4^m}
          {z^m \over \lambda^{2 m} } \right) =
   h(\lambda) + 2 {N \over \beta} + O(\lambda^{-1}).
\end{equation}

\noindent
The $O(\lambda^0)$ terms determine $z$:

\begin{equation}
\sum_{m \geq 1} c_m {(2 m)! \over m!^2 4^m} z^m =
     2 {N \over \beta}.                                     \label{zd1}
\end{equation}

We can express the partition function ${\cal Z}$
as a power series in the couplings
$c_m$ and the parameter $N/\beta$.  If we consider the potential $V$ and
hence the $c_m$ to be fixed, then ${\cal Z}$
has a radius of convergence in
$N/\beta$; we construct the potential so that this radius of convergence
is $1$.  For $N/\beta > 1$---that is,
for (renormalized) temperatures higher than a critical temperature---the
series expansion of ${\cal Z}$ diverges.
The behavior of ${\cal Z}$ near the critical
point is what describes the quantum geometry of random surfaces, since
at the critical point the behavior of ${\cal Z}$ is dominated by Feynman graphs
of large order.  It turns out that the singular behavior of ${\cal Z}$ at the
critical point is determined by the dependence of $z$ on $N/\beta$ in
\equno{zd1}.  The \kth multicritical model is constructed by finding $c_m$
such that \equno{zd1} takes the form

\begin{equation}
1 - \left( 1 - {z \over z_c} \right)^k = {N \over \beta} \equiv
     1 - \mu_0
\end{equation}

\noindent
for some $z_c$.  We shall use

\begin{equation}
c_m = \left\{
   \begin{array}{ll}
   \displaystyle{(-1)^{m+1} {2 k!\, m! \over (2m)!\, (k-m)!\, k^m}} \qquad &
         {\rm for} \quad m \leq k   \\
    \displaystyle{0} \qquad & {\rm for} \quad m > k
   \end{array} \right.
\end{equation}

\noindent
which gives $z_c = 4 k$.

Let $\epsilon$ be the lattice spacing of our random
lattice.  Then area is measured in units of $\epsilon^2$, which will
be taken to zero as the average number of plaquettes in
the discretizations of the random surface diverges, so as to yield
finite area.  The parameter $\mu_0$ corresponds to the lowest dimension
operator in the \kth multicritical theory, and it was at first believed that
if one wrote $\mu_0 = (2 \epsilon)^k t_0$,
then $t_0$ would be the cosmological constant.  This is false for
the $k>2$ models, which are non-unitary:
$t_0$ has dimensions of $({\rm Length})^{-k}$, and is the coupling
constant for the gravitationally dressed conformal
field of the lowest dimension.
The true cosmological constant is by definition conjugate to area,
so it has dimensions of $({\rm Length})^{-2}$.

We must consider a more general
perturbation around criticality.  Specifically, we perturb the potential
by replacing $c_m \to c_m + {2 \, m!^2 \over (2 m)! k^m} \mu_m$ for
all $m>0$.  \equno{zd1} then reduces to

\begin{equation}
\left( 1 - {z \over z_c} \right)^k =
     \sum_{m \geq 0} \mu_m \left( {z \over z_c} \right)^m.
\end{equation}

\noindent
Higher dimension scaling operators are introduced by choosing the $\mu_m$
so the that right hand side becomes $t_l \epsilon^{k-l} (1 - z/z_c)^l$
for some $l < k$, where $t_l$ is the coupling to the $l^{\rm th}$
scaling operator.  As was shown in \cite{MSS}, the couplings to the
operators corresponding to conformal fields in the Liouville theory
are analytical combinations of $t_l$ with a definite overall dimension.
The cosmological constant $\tau$ corresponds to a perturbation with
$t_l = \alpha_l 2^{k-l} \tau^{(k-l)/2}$ for $l-k$ even and
$0 \leq l \leq k-2$.  The $\alpha_l$ are
real numbers chosen so that the singular part
of the one-loop amplitude obeys a Bessel equation, which emerges from the
Wheeler de Witt equation.

\begin{equation}
\left( 1 - {z \over z_c} \right)^k = \sum_{l} \alpha_l
   (2 \epsilon \sqrt{\tau})^{k-l} \left( 1 - {z \over z_c} \right)^l\ ,
                                                          \label{CosmCnst}
\end{equation}

\noindent
and by requiring $\sum_l \alpha_l = 1$ (which is equivalent to fixing a
normalization for $\tau$) we find $z = z_c (1 - 2 \epsilon \sqrt{\tau})$.
For $k=2$ and $k=3$ there is only one $\alpha_l$, so $\tau \propto t_{k-2}$.
For $k=4$, it is found that $\alpha_0 = -1/5$ and $\alpha_2 = 6/5$.  We
do not know a general way to determine the $\alpha_l$ except by straight
calculation.

The partition function for surfaces with one boundary is the continuum
limit of the Green's function of the field theory above.
The $l$-point Green's function is

\begin{equation}
G_l = \langle \Tr \Phi^l \rangle =
     \int_{-\sqrt{z}}^{\sqrt{z}} d\lambda \, u(\lambda) \lambda^l.
\end{equation}

\noindent
It is convenient to introduce $G(y) = \sum_{l \geq 0} G_l\, y^l$, since
it is easy to see that

\begin{equation}
G(y) = {\beta \over N} {1 \over 2 y} F\left( {1/ y} \right) =
     {\beta \over N} {1 \over 2} \left( g\left( {1/ y} \right) -
       h\left( {1/ y} \right) \sqrt{1 - z y^2} \right).
\end{equation}

\noindent
Since at $z = z_c$ the radius of convergence of $G(y)$ is
$y_c = 1/\sqrt{z_c}$, the continuum limit is taken by setting
$y = y_c \exp (-\epsilon \zeta)$, where now $\zeta$ is conjugate to the
length of the boundary.  The continuum limit ($\epsilon \to 0$) of
$G(y)$ was calculated using Mathematica, with the following results:

\begin{equation}
\begin{array}{ll}
k=2: \  & G(\zeta) = {4 \over 3} - {8 \over 3} \zeta \epsilon +
       {16 \sqrt{2} \over 3} \epsilon^{3/2} f_2(\zeta,\tau) +
       O(\epsilon^2) \\
    & \qquad\quad {\rm where} \quad f_2(\zeta,\tau) =
        2 \left( \zeta - {1 \over 2} \sqrt{\tau} \right)
         \sqrt{\zeta + \sqrt{\tau}} \\
k=3: \  & G(\zeta) = {6 \over 5} - {4 \over 5} \zeta \epsilon +
       {4 (-4 \tau + 7 \zeta^2) \over 5} \epsilon^2 +
       {96 \sqrt{2} \over 5} \epsilon^{5/2} f_3(\zeta,\tau) +
       O(\epsilon^3) \\
    & \qquad\quad {\rm where} \quad
        f_3(\zeta,\tau) = {2\over 3}\left( -\zeta^2 +
          {1 \over 2} \zeta \sqrt{\tau} + {1 \over 4} \tau \right)
         \sqrt{\zeta + \sqrt{\tau}} \\
k=4: \  & G(\zeta) = {8 \over 7} - {16 \over 35} \zeta \epsilon +
       {16 (-2 \tau + 3 \zeta^2) \over 35} \epsilon^2 +
       {32 \zeta (174 \tau - 185 \zeta^2) \over 525} \epsilon^3 +
       {512 \sqrt{2} \over 7} \epsilon^{7/2} f_4(\zeta,\tau) +
       O(\epsilon^4) \\
    & \qquad\quad {\rm where} \quad
        f_4(\zeta,\tau) = {2\over 5}\left( \zeta^3 -
          {1 \over 2} \zeta^2 \sqrt{\tau} - {1 \over 2} \zeta \tau +
          {1 \over 8} \tau^{3/2} \right)
         \sqrt{\zeta + \sqrt{\tau}}
\end{array}                                                     \label{ContG1}
\end{equation}

\noindent
The coefficient $f_k(\zeta,\tau)$ of the leading nonanalytic term in
$\epsilon$ is the
universal part of the disk amplitude.  $f_k(\zeta,\tau)$ is, up to a
numerical factor, the Laplace transform of the disk amplitude
$F_k(L,\tau)$ in \equno{ExactFL}.  The lower order analytic terms in
$\zeta$ and $\tau$ correspond to zero length and zero area terms, and most
of them can be dropped, for the following reason.  We are free to adjust
$\zeta$ by an analytic function of $\zeta$ and $\tau$, which, in order to
avoid trivial additive and multiplicative rescalings of $\zeta$, and
to preserve the dimension of $\zeta$, should have the form

\begin{equation}
\zeta \to \zeta + r_1 \epsilon \tau + r_2 \epsilon \zeta^2 +
   r_3 \epsilon^2 \tau \zeta + r_4 \epsilon^2 \zeta^3 + \ldots
                                                              \label{Redef}
\end{equation}

\noindent
where the $r_i$ are $c$-numbers.
Such a redefinition preserves the universal term $f_k(\zeta,\tau)$ and
corresponds simply to a different way of treating zero area and zero length
terms, as pointed out in \cite{MSS}.  By an appropriate choice of the
$r_i$ in \equno{Redef}, we can absorb all but the first two analytic
terms of the expressions in \equno{ContG1} into $\zeta$.  We then find
the simple form

\begin{equation}
G(\zeta) = {2k \over 2k-1} \left( 1 - {\zeta \epsilon\over \sigma} \right) +
      \alpha \epsilon^{k-1/2} f_k(\zeta,\tau) + O(\epsilon^k)  \label{ContG2}
\end{equation}

\noindent
where

\begin{equation}
\sigma = k - 3/2
\end{equation}

\noindent
and $\alpha$ is a numerical factor chosen for
each $k$ so that the leading term of a small $\tau$ expansion of
$f_k(\zeta,\tau)$ is $(-1)^k \zeta^{\sigma +1}/\sigma$.

For later convenience, we mention one more mathematical point: since
$z/z_c = 1 + O(\epsilon)$, we can alter the perturbation \equno{CosmCnst}
to read

\begin{equation}
\left( 1 - {z \over z_c} \right)^k = \sum_{l} \alpha_l
   (2 \epsilon \sqrt{\tau})^{k-l} \left( 1 - {z \over z_c} \right)^l
   \left( z \over z_c \right)^n
\end{equation}

\noindent
where $n$ is a fixed integer, and the equation
$z = z_c (1 - 2 \epsilon \sqrt{\tau})$ will receive corrections that are
analytic in $\epsilon$ and of order $\epsilon^2$ and higher.  The
non-universal terms in \equno{ContG1} will change, but the leading analytic
and leading nonanalytic terms will be unaffected, so by making an appropriate
redefinition of the form \equno{Redef}, we still arrive at \equno{ContG2}.
If we take $n=2$, then $\mu_0 = 0$ and $\mu_1 = 0$.
This is desirable because it means that the weights of planar Feynman
diagrams do not depend on the number of edges.

\section{The transfer matrix}
\label{TransferM}

The focus of \cite{Kawai1} is the
analysis of the evolution of a loop through some fixed geodesic
distance on the surface, where on a discretized surface geodesic distance
is defined as the minimal number of plaquettes one must traverse to get
from one point to another.  The goal is to calculate the partition function
of a tube with one entrance loop and one exit loop, such that each
point on the exit loop is a fixed geodesic distance $D$ from the entrance
loop.

On a discretized surface, one starts with an
entrance loop $\gamma$ (see Fig.~1),
defines a ``forward'' direction for geodesic distance (inward in
Fig.~1), and thinks of advancing the loop along the lattice
one step at a time.  To accomplish this ``one-step deformation,'' as it
was called in \cite{Kawai1}, one first removes any double links
that may exist on $\gamma$ and then moves each remaining link across the
plaquette it borders in the forward direction.  Clearly, this process
will sometimes split $\gamma$ into several loops.  To phrase it another way,
once we have removed all double links from $\gamma$, we color all the
plaquettes that the remaining links border in the forward direction.  Each
side of a colored plaquette that is not part of the entrance loop $\gamma$
becomes a link in one of the exit loops.  Since we are interested
in having just one exit loop, we designate one of the loops evolved from
$\gamma$ as {\it the} exit loop, and close off the other loops with disks
generated by the disk amplitude of the matrix model.  A series of
one-step deformations gives an exit loop which is a fixed geodesic
distance from the entrance loop.

We now want to ask the following combinatorical question: given that the
entrance loop $\gamma$ has $l$ links and the exit loop $\gamma'$ has
$l'$ links, how many ways are there to construct a discretization between
$\gamma$ and $\gamma'$, using a given number
of squares, hexagons, octagons, etc.,
such that $\gamma'$ is the exit loop evolved from $\gamma$ in the
course of a one-step deformation, as defined above?  Or, in terms of
the Feynman diagram which is the dual graph of the discretization, how
many planar diagrams can be drawn on the
surface of a cylinder with $l$
external legs pointing ``down'' and $l'$ external legs pointing ``up'',
with specified numbers of vertices of order four, six, eight, etc., and
such that every ``up'' leg is connected to the same vertex as
some ``down'' leg (but not necessarily vice versa)?
To answer this question, we assign a multiplicative weight $g_m$ to each
$2m$-gon (equivalently, to each vertex of order $2m$ on the dual lattice),
assign as an overall weight to each of the discretizations between
$\gamma$ and $\gamma'$ the product of the weights of the plaquettes
used to build it, and let $N_{l,l'}$ be the sum of the weights of all
such discretizations.  $N_{l,l'}$ is the generating function that answers
the above combinatorical question, but of course its more interesting
property is that it is the partition function of tubes, or rather ribbons,
of geodesic width one (in lattice units) and with entrance and exit loops
of $l$ and $l'$ links, respectively.

We propose to evaluate

\begin{equation}
N(y,y') = \sum_{l,l' \geq 1} N_{l,l'} \, y^l (y')^{l'}
\end{equation}

\noindent
using combinatorics and the disk amplitude from the matrix models.
For calculational convenience we will designate the entrance loop as
unmarked and the exit loop as marked.  On the dual lattice, this corresponds
to considering the $l$ external legs of the planar diagram that
point ``down'' to be distinguishable only up to cyclic permutations and
the $l'$ external legs that
point ``up'' to be completely distinguishable.  A different
convention of marking would change $N_{l,l'}$ only by a factor of $l$ or
$l'$.  The Green's function $G(y)$ of the matrix field theory refers to
diagrams in which the external legs are distinguishable---that is, $G(y)$
is the amplitude of a disk with a marked boundary.  The weight $g_m$
assigned to a $2m$-gon by the Feynman rules for the matrix field theory is

\begin{equation}
g_m = -c_m - {2 \, m!^2 \over (2 m)! k^m} \mu_m \ .
\end{equation}

\noindent
Let us first consider the case where only squares are used in the
discretizations.
All the discretizations contributing to $N(y,y')$
may be built up as follows (here we are paralleling closely the work
of \cite{Kawai1}, only using squares instead of triangles).  The marked
point on the exit loop must border one of the shapes in list a) of
Fig.~2: the curved loops on the last three shapes denote the insertion
of a disk.  As we follow the exit loop around its length, we can encounter
any of the shapes in list b) in any order: the entrance and exit loops
are connected from the corner of one shape to the next, and in the end
the last shape is connected back to the original member of list a), closing
the loop.

It is perhaps not transparent why the weighting of each shape
is what it is, so let us work through the example shown in Fig.~3.
The loop above the square indicates the insertion of a disk amplitude,
which must be marked because something must pick out the point where the
disk boundary meets the square.  The disk discretizations with less than
two sides on the boundary are omitted because there have to be two
boundary sides that match with the two free sides of the square.
Let us think now in terms of the dual lattice.
Given any Feynman graph contributing to the disk amplitude, we
form a graph of the desired type by attaching to an adjacent pair of its
external legs another vertex.  The weight of the original disk amplitude
graph must be multiplied by $g_2$ to get the weight of the new graph,
because we have added one vertex and tied up two external legs while
adding two new external legs.  Summing over all allowable disk amplitude
graphs (those with at least two external legs) then gives a total weight
of $g_2 (G(y) - G_0 - G_1 y)$, as
claimed.  Incidentally, the first, ``undrawable'' term $g_2 y^2$ in
Fig.~3 corresponds to the two free sides of the square being
identified.  It might be helpful for the reader to identify which disk
amplitude diagrams correspond to each term in Fig.~3.

Now it is not hard to write down the weight that would be assigned to a
$2m$-gon bordered on $n$ sides by the exit loop.  Starting with a disk
amplitude graph with at least $2m-n-2$ external legs, we add to it one
$2m$-vertex, tying up $2m-n-2$ of its external legs and adding $n+2$
new external legs, $2$ of which are entrance legs and $n$ of which are
exit legs.  Thus the total weight is

\begin{equation}
y'^n y^{n+4-2m} g_m \left( G(y) - \sum_{l=0}^{2m-n-3} G_l y^l \right).
\end{equation}

\noindent
Note that because of the planarity of the surface, the sides of the
$2m$-gon bordered by the exit loop must be contiguous.  When $2m$-gons
are allowed in the discretization, we include all the $2m$-gon shapes
in list b), and also in list a) with a multiplicity determined by the
number of exit links on the shape.

The combinatorical problem is now quite simple: we construct an arbitrary
discretization from one member of list~a) and some sequence of members of
list~b).  The total amplitude when only squares are used is

\begin{eqnarray}
N(y,y') &=& \left( 3 y'^3 y g_2 + 2 y'^2 y^2 g_2 G(y) +
              y' y g_2 (G(y) - G_0) \right) \times   \nonumber \\
        & & \sum_{n=0}^{\infty} [y'^3 y g_2 +
                         y'^2 y^2 g_2 G(y) +
                         y' y g_2 (G(y) - G_0) +
        g_2 (G(y) - G_0 - G_1 y) +
                         y^2 G(y) ]^n                \nonumber
\end{eqnarray}

\begin{equation}
      = {3 y'^3 y g_2 + 2 y'^2 y^2 g_2 G(y) +
               y' y g_2 (G(y) - G_0) \over
        1 - y'^3 y g_2 - y'^2 y^2 g_2 G(y) -
                y' y g_2 (G(y) - G_0) -
        g_2 (G(y) - G_0 - G_1 y) - y^2 G(y) }\ . \label{N1k4}
\end{equation}
\noindent
Expanding out the first expression gives terms corresponding to individual
discretizations.  Let us define $S$ to be $1$ minus the sum of weights in
list~b):

\begin{equation}
S = 1 - y'^3 y g_2 - y'^2 y^2 g_2 G(y) - y' y g_2 (G(y) - G_0) -
          g_2 (G(y) - G_0 - G_1 y) - y^2 G(y).
\end{equation}

\noindent
Recalling that each shape from list~b) enters into list~a) with a
multiplicity equal to the number of exit links on the shape, we find that

\begin{equation}
N(y,y') = -{\partial \log S \over \partial \log y'}\ .           \label{NS}
\end{equation}

\noindent
In the general case where arbitrary $2m$-gons are allowed, list~a) and
list~b) are appropriately expanded, and

\begin{eqnarray}
S &=& 1 - y^2 G(y) - \sum_{m \geq 1} \left( y'^{2m-1} y g_m +
        \sum_{n=0}^{2m-1} y'^n y^{n+4-2m} g_m \left( G(y) -
          \sum_{l=0}^{2m-n-3} G_l y^l \right) \right)    \nonumber \\
  &=& 1 - y y' + {y \over y'} g(y') - 2 y^2 G(y) -
        y^4 \sum_{m \geq 1} {g_m \over y^{2 m}}
          \sum_{n=0}^{2m-2} (y' y)^n \left( G(y) -
            \sum_{l=0}^{2m-n-3} G_l y^l \right),               \label{GenS}
\end{eqnarray}

\noindent
and \equno{NS} still holds.

The continuum limit of $N(y,y')$ is taken by expanding about the radius
of convergence of each of the variables.  The convergence of $S$ is
determined by the convergence of $G(y)$, which we explained in
Section~\ref{MMloops}: $y_c = 1/\sqrt{z_c}$, and we set
$y = y_c \exp (-\epsilon \zeta)$.  For fixed $y \leq y_c$, $S$ is entire in
$y'$, but $N(y,y')$ is analytic in $y'$ only up to the magnitude of the zero
of $S$ nearest $y' = 0$.  It turns out that for $\zeta = 0$ and $\tau = 0$,
this zero is at $y' = y_c^{-1}$.  Hence we set
$y' = y_c^{-1} \exp (-\epsilon \zeta')$.  An analytic redefinition of
$\zeta'$ of the form

\begin{equation}
\zeta' \to \zeta' + r_1 \epsilon \tau + r_2 \epsilon \zeta^2 +
     r_3 \epsilon \zeta \zeta' + r_4 \epsilon \zeta'^2 + \ldots
\end{equation}

\noindent
is allowed, for the same reasons as for the redefinition \equno{Redef}.
Thus when we expand $N(y,y')$ in $\epsilon$, we need only retain the leading
analytic and leading nonanalytic terms in $\zeta'$.
Writing $N(\zeta,\zeta')$ in place of $N\left( y_c \exp (-\epsilon \zeta),
y_c^{-1}\exp (-\epsilon \zeta') \right)$, we find remarkably simple
results:

\begin{equation}
N(\zeta,\zeta') = {1 \over \epsilon} \left({1\over \zeta + \zeta' -
      \alpha' \epsilon^{\sigma} f(\zeta,\tau)} +
      O(\epsilon^{k-1}) \right)                                \label{ContN}
\end{equation}

\noindent
where $\alpha'$ is another numerical factor.

Let $N_{l,l'}(d)$ be the lattice
partition function of tubes of geodesic length
$d$ with an unmarked entrance loop of $l$ links and a marked exit loop of
$l'$ links.  The great insight of \cite{Kawai1} is that this object has a
simple composition law:

\begin{equation}
N_{l,l'} (d_1 + d_2) = \sum_{l'' = 1}^{\infty}
   N_{l,l''} (d_1) N_{l'',l'} (d_2)\ .                         \label{Comp1}
\end{equation}

\noindent
In rough terms, we can cut a tube in two at some geodesic time and find
its amplitude by summing the products of the amplitudes of the pieces
over all possible lengths of the intermediate loop (see Fig.~4).
The motivation for the convention of taking the entrance loop to be unmarked
and the exit loop to be marked lies in the fact that there are
$l''$ ways to glue the exit loop of one tube to
the entrance loop of another---$l''$ being the number of links on each
loop---but we want to avoid factors of $l''$ in the composition
law.  Suppose we mark the entrance loop of the rightmost
tube in Fig.~4; such tubes would have amplitude
$l'' N_{l'',l'} (d_2)$.  We now can glue the two tubes
together in such a way that the two marks are at any of $l''$ positions
relative to each other, so the amplitude for the resulting surface would be
$l'' \left( N_{l,l''} (d_1) l'' N_{l'',l} (d_2) \right)$.  But that surface
would still have the two marks on the intermediate loop, and we must delete
them to get a surface of the type shown on the left side of Fig.~4.
Thus we divide our last expression by $l''^2$ to get
$N_{l,l''} (d_1) N_{l'',l'} (d_2)$ and sum over $l''$ to get \equno{Comp1}.

Since \equno{Comp1} is just a matrix product,
it is clear that $N_{l,l'} (d)$ is
the time evolution kernel of some Hamiltonian.  We will
find the continuum limit of this Hamiltonian and then calculate
$N_{l,l'} (d)$ by solving the corresponding Schr\"odinger equation.

In terms of $N(y,y',d) =
\sum_{l,l' \geq 1} N_{l,l'} (d) \, y^l (y')^{l'}$, \equno{Comp1}
takes the form

\begin{equation}
N(y,y',d_1 + d_2) = {1 \over 2 \pi i } \oint {dx \over x}
   N(y,x,d_1) N(1/x,y',d_2)
\end{equation}

\noindent
where the integral is taken along a contour around the origin.
Making the change of variables $x = y_c^{-1} \exp (-\epsilon \xi)$, we find

\begin{equation}
N(\zeta,\zeta',d_1 + d_2) = {\epsilon \over 2 \pi i}
   \int_{-i \pi / \epsilon}^{i \pi / \epsilon} d\xi
     N(\zeta,\xi,d_1) N(-\xi,\zeta',d_2) \to
   {\epsilon \over 2 \pi i} \int_{-i \infty}^{i \infty} d\xi
     N(\zeta,\xi,d_1) N(-\xi,\zeta',d_2)                       \label{Comp2}
\end{equation}

\noindent
for small $\epsilon$.  In particular, to the first nontrivial order in
$\epsilon$,

\begin{eqnarray*}
N(\zeta,\zeta',d+1) &=& {\epsilon \over 2 \pi i}
      \int_{-i \infty}^{i \infty} d\xi
        N(\zeta,\xi) N(-\xi,\zeta',d) \\
  &=& {1 \over 2 \pi i} \int_{-i \infty}^{i \infty} d\xi
        {1 \over \zeta + \xi -
           \alpha' \epsilon^{\sigma} f(\zeta,\tau)}
        N(-\xi,\zeta,d) \\
  &=& N\left( \zeta - \alpha' \epsilon^{\sigma} f(\zeta,\tau),
        \zeta',d \right) =
    N(\zeta,\zeta',d) - \alpha' \epsilon^{\sigma} f(\zeta,\tau)
      {\partial \over \partial \zeta} N(\zeta,\zeta',d)\ ,
\end{eqnarray*}

\noindent
where we have closed the contour to the left because
the function $N(-\xi,\zeta,d)$ has a branch cut for positive real
$\xi > \sqrt{\tau}$.  Setting $D = \alpha' \epsilon^{\sigma} d$ and
taking $\epsilon \to 0$,
we find that $N(\zeta,\zeta',D)$ is a solution to the equation

\begin{equation}
{\partial \over \partial D} \psi(\zeta,D) =
  -f(\zeta,\tau) {\partial \over \partial \zeta} \psi(\zeta,D) \label{DiffEQ}
\end{equation}

\noindent
with the initial condition read off from \equno{ContN},
$N(\zeta,\zeta',0) = 1/(\zeta + \zeta')$.\footnote{In order
to work with finite quantities in the continuum limit, we rescale
$N(\zeta,\zeta',D)$ to absorb the factor $1/\epsilon$ present in the
lattice definition.}
In \cite{Kawai1}, this equation was derived for the case $k=2$, but using
only triangles in the discretizations, rather than squares; thus for this
case we have a universality check.  What is interesting is that the same
equation holds for higher $k$, with the disk amplitude $f(\zeta,\tau)$
adjusted appropriately, and with a new scaling behavior for macroscopic
geodesic distance $D$.  The definition $D = \alpha' \epsilon^{\sigma} d$
is very interesting,
and in a way unexpected, because it means that $D$ has the dimension of
$L^{\sigma}$ instead of just $L$ (recall $\sigma = k - 3/2$).

In what follows it is convenient to introduce $N(\zeta,L',D)$, the
inverse Laplace transform of $N(\zeta,\zeta',D)$ with respect to
$\zeta'$.
Since \equno{DiffEQ} involves only $\zeta$ and $D$,
$N(\zeta,L',D)$ is also a solution of this differential equation, only with
different initial conditions, namely $N(\zeta,L',0) = \exp(-\zeta L')$.
Our next step will be to find an approximate
method for solving the differential equation and to use the solutions, as
was done in \cite{Kawai1}, to determine universal functions describing
the way random surfaces branch.
These functions directly depend on $N(L,L',D)$,
the inverse Laplace transform of $N(\zeta,L',D)$ with respect to
$\zeta$.  $N(L,L',D)$ is
the continuum amplitude for tubes of length $D$ with boundaries
of length $L$ and $L'$, and it satisfies the initial condition
$N(L,L',0) = \delta(L-L')$.

\section{Calculating the tube amplitude}
\label{TubeAmp}

A solution to \equno{DiffEQ} may be found by the method of characteristic
curves: if the function $\zeta_0 (\zeta,D)$ solves the ODE

\begin{equation}
{d \over d D} \zeta_0(\zeta,D) = -f(\zeta_0(\zeta,D),\tau)\ ,
   \qquad \zeta_0(\zeta,0) = \zeta                            \label{CCEQ1}
\end{equation}

\noindent
then $\psi(\zeta,D) = \psi(\zeta_0(\zeta,D),0)$ is a solution to
\equno{DiffEQ}.  Geometrically speaking,
$\psi(\zeta,D)$ is constant along a family
of (non-intersecting) characteristic curves in $\zeta$-$D$ space;
$\zeta_0(\zeta,D)$ is the $\zeta$-coordinate of the point where the
characteristic curve passing through $(\zeta,D)$ meets the line $D = 0$.
{}From \equno{CCEQ1} it follows that $\zeta_0(\zeta,D)$ is implicitly
determined by

\begin{equation}
\int_{\zeta_0}^{\zeta} d\xi {1\over f(\xi,\tau)} = D\ .   \label{IntEQ}
\end{equation}

\noindent
Now $N(\zeta, L',D)$, the solution of \equno{DiffEQ} with initial conditions
$N(\zeta,L',0) = \exp (-\zeta L')$, is simply given by

\begin{equation}
N(\zeta,L',D) = \exp \left( -\zeta_0 (\zeta,D) L' \right)\ .   \label{DSol}
\end{equation}

\noindent
The inverse Laplace transform of this function gives the desired tube
amplitude $N(L, L', D)$.
Thus, the problem is reduced to the explicit determination
of $\zeta_0(\zeta,D)$.

For $k=2$, $\zeta_0(\zeta,D)$ can be expressed in terms of
elementary functions. Indeed, after evaluating the integral in \equno{IntEQ}
with $f(\zeta,\tau) = \left(2 \zeta - \sqrt{\tau} \right)
\sqrt{\zeta + \sqrt{\tau}} $, and some algebra, we arrive at

\begin{equation}
\zeta_0 = -\sqrt\tau +{3\over 2}\sqrt\tau \coth^2
\left (\sqrt{3\over 2}\tau^{1/4} D-{1\over 2}\log
{\sqrt{\zeta+\sqrt\tau}-\sqrt{3\over 2}\tau^{1/4}\over
\sqrt{\zeta+\sqrt\tau}+\sqrt{3\over 2}\tau^{1/4} }\right ) \label{Exact}
\end{equation}

\noindent
Unfortunately, it seems impossible to find the exact inverse
Laplace transform of \equno{DSol}. Instead, we will look for the
behavior of $N(L,L',D)$ in the limit of a point-like entrance loop
($L\to 0$), as well as in the limit of a very large entrance loop
($L\to\infty $). In the first case, it is sufficient to study the
dominant behavior of \equno{Exact} for large $\zeta$, and in the
second case, for small $\zeta$.

Let us first consider the limit of a point-like entrance loop.
Substituting \equno{Exact} into \equno{DSol} and expanding for
large $\zeta$, we find

\begin{equation}
N(\zeta,L',D) = e^{L'\sqrt\tau} e^{-{3\over 2}L'\sqrt\tau \coth^2
\left (\sqrt{3\over 2}\tau^{1/4} D\right )}+
{1\over\sqrt \zeta} e^{L'\sqrt\tau} {\partial\over\partial D}
e^{-{3\over 2}L'\sqrt\tau
\coth^2\left (\sqrt{3\over 2}\tau^{1/4} D\right)}+ O(1/\zeta)
\label{Largez}
\end{equation}

\noindent
Performing the inverse Laplace transform, we find

\begin{equation}
N(L\to 0, L',D) =
{1\over\sqrt {\pi L}} e^{L'\sqrt\tau} {\partial\over\partial D}
e^{-{3\over 2}L'\sqrt\tau
\coth^2\left (\sqrt{3\over 2}\tau^{1/4} D\right)}+ O(L^0)
\label{SmallL}
\end{equation}

\noindent
This formula is valid for $L\ll D^2, ~\tau^{-1/2}$. Nothing is
assumed, however, about the relative magnitudes of $D^2$ and
$\tau^{-1/2}$. In \cite{Kawai1} the calculation was performed in the
regime
$L \ll D^2 \ll \tau^{-1/2}$
so that the total area of the surface could
be sent to infinity while $D$ and $L'$ were kept finite.
Our \equno{SmallL} generalizes the result of \cite{Kawai1} and reduces
to it when expanded for small $\tau$. For $k>3$, however, we cannot
find an explicit form of $\zeta_0(\zeta,D)$ analogous to
\equno{Exact}. For that reason we will work in the limit
$1/\zeta \ll D^{1/\sigma} \ll \tau^{-1/2}$ (recall $\sigma = k - 3/2$),
and expand $\zeta_0$ in powers of $1/\zeta$ and $\tau$.

Since $f_k(\xi,\tau) = (-1)^k~ {\xi^{\sigma+1}\over \sigma}
(1 + O(\tau / \xi^2))$,
as our first approximation we have from \equno{IntEQ},

\begin{equation}
\zeta_0 (\zeta,D) = \left( (-1)^k D +
    \zeta^{-\sigma} \right)^{-1/\sigma} + O(\tau)  \ .       \label{SickApp}
\end{equation}

\noindent
{}From this formula we can see a
sickness in the case of odd $k$: as $D \to \zeta^{-\sigma}$ from below,
$\zeta_0 (\zeta,D) \to \infty$.  What is happening is that the solution
to \equno{CCEQ1} with $\tau = 0$ is running off to $\infty$ in finite
time. \equno{SickApp} becomes complex for
$D > \zeta^{-\sigma}$, but the real solution
to \equno{CCEQ1} just stays at $\infty$.  Letting
$\tau$ become finite does not help the situation except when the initial
$\zeta$ is less than the rightmost zero of $f(\zeta,\tau)$,
which is at $\zeta\sim\sqrt{\tau}$. Since $N(\zeta, L', D)$ vanishes
for large enough real $\zeta$, the inverse Laplace transform
$N(L,L',D)$ cannot be positive definite for small $L$. This seems to
lead to meaningless results, and we will not consider the
limit of a point-like entrance loop in the odd $k$ models.  It may be
that if we worked with fixed $L$ from the beginning, instead of with
Laplace-tranformed equations like \equno{DiffEQ}, we could avoid the
sickness we have observed.

The even $k$ models avoid
this sickness,  as is clear from
Fig.~5 where we show the qualitative behavior of
$\zeta_0 (\zeta,D)$ for $k=3$ and $k=4$ and finite $\tau$.
To determine the baby universe distribution function in the case of even
$k$, we need to continue the expansion of \equno{SickApp} in powers of
$\tau$ and find the coefficient of the leading fractional power.  First
we expand

\begin{equation}
{1\over f_k(\xi, \tau)}={k-3/2\over\xi^{k-1/2}}
+\sum_{n=1}^{k-1} a_n\tau^n {k+2n-3/2\over\xi^{k+2n-1/2}}
+b\tau^{k-1/2}{3k-5/2\over\xi^{3k-3/2}} +O(\tau^{k})\ ,
                                                           \label{App1}
\end{equation}

\noindent
where $a_n$ and $b$ are numerical coefficients which can be
found from \equno{ContG1}. Substituting
\equno{App1} into \equno{IntEQ} and integrating term-by-term,
we obtain the following relation:

\begin{eqnarray}
\zeta_0 (\zeta,D)& = \left( D +
    \zeta^{-\sigma} \right)^{-1/\sigma} +\tau{a_1\over\sigma}
\left( D +\zeta^{-\sigma} \right)^{-1-1/\sigma}\left (\zeta_0^{-\sigma-2}
-\zeta^{-\sigma-2}\right )+\ldots+O(\tau^{k-1}) \cr &
+\tau^{k-1/2}~ {b\over\sigma}
\left( D +\zeta^{-\sigma} \right)^{-1-1/\sigma}
\left (\zeta_0^{-3\sigma-2}-\zeta^{-3\sigma-2}\right )+O(\tau^k) \ .
\label{ItApp}
\end{eqnarray}

\noindent
Now the expansion of $\zeta_0$ in powers of $\tau$ can be found
iteratively. The coefficient of the leading fractional power,
$\tau^{k-1/2}$, is actually obtained after one iteration.  Thus, the
desired expansion has the form

\begin{eqnarray}
\zeta_0 (\zeta,D) &= \left( D +
    \zeta^{-\sigma} \right)^{-1/\sigma} +\tau{a_1\over\sigma}
\biggl [\left( D +\zeta^{-\sigma} \right)^{1/\sigma}-\zeta^{-\sigma-2}
\left( D +\zeta^{-\sigma} \right)^{-1-1/\sigma}\biggr ]
+\ldots+O(\tau^{k-1}) \cr &
+\tau^{k-1/2}~ {b\over\sigma}
\biggl [\left( D +\zeta^{-\sigma} \right)^{2+1/\sigma}-\zeta^{-3\sigma-2}
\left( D +\zeta^{-\sigma} \right)^{-1-1/\sigma}\biggr ]+O(\tau^k) \ .
\label{CompExp}
\end{eqnarray}

\noindent
Substituting this into \equno{DSol}, we have

\begin{eqnarray}
&N(\zeta,L',D) = e^{- L'\left( D +
    \zeta^{-\sigma} \right)^{-1/\sigma}}\biggl\{ 1 -\tau L'{a_1\over\sigma}
\biggl [\left( D +\zeta^{-\sigma} \right)^{1/\sigma}-\zeta^{-\sigma-2}
\left( D +\zeta^{-\sigma} \right)^{-1-1/\sigma}\biggr ]
+\ldots\cr & +O(\tau^{k-1})
-\tau^{k-1/2} L'{b\over\sigma}
\biggl [\left( D +\zeta^{-\sigma} \right)^{2+1/\sigma}-\zeta^{-3\sigma-2}
\left( D +\zeta^{-\sigma} \right)^{-1-1/\sigma}\biggr ]+O(\tau^k) \biggr\}
\ .\label{CompTrans}
\end{eqnarray}

\noindent
Now we expand this for large $\zeta$ and inverse Laplace transform
term-by-term, discarding the $O(\zeta^0)$ pieces, which give zero
length terms.  We find

\begin{eqnarray}
N(L\to 0,L',D) &=&
     {L^{\sigma - 1} \over \Gamma (\sigma)} e^{-x}
   \biggl( {L' \over \sigma D^{1+1/\sigma}} +
      O(\tau) + \ldots+O(\tau^{k-1}) +              \nonumber \\
  & & \gamma_1 D L' ((2k-2) D^{1/\sigma} +
        L') \tau^{k-1/2} + O(\tau^k)\biggr) +
   O(L^{2 \sigma-1})                                           \label{NLL}
\end{eqnarray}

\noindent
where $x = L'/D^{1/\sigma}$ and $\gamma_1=-b/\sigma^2$.

\section{The branching structure of random surfaces}
\label{BStructure}

We now wish to extract from $N(L,L',D)$ a universal function which describes
the branching structure of a large surface of planar topology and one
boundary loop of length $L_0$.  We will
find that such surfaces have lots of little protuberances---hair, if you
will.  Let $R(D)$ be the part of the surface which is at a
geodesic distance $D$ or less from the boundary loop.  Let $\rho(L_0, L,D)$
be the distribution of boundary loop lengths for $R(D)$: on average,
$R(D)$ has $\rho(L_0, L,D) dL$ boundary loops with length in the interval
$(L,L+dL)$.  This quantity was introduced and calculated for $k=2$ and
$L_0\to 0$ in \cite{Kawai1}.  The strategy for calculating it is as follows.
The partition function for all surfaces is the disk amplitude $F(L_0)/L_0$.
Let us think for a moment of the discrete version $\rho(l_0, l,d)$ of
$\rho(L_0, L,D)$.  This
function is the statistical average over all disks with boundary length
$l_0$ of the number of loops with $l$ links at geodesic distance
$d$ from the boundary.  According to \cite{Kawai1}, in the limit of
large disk area

\begin{equation}
\rho(l_0,l,d) = \lim_{\tau\to 0}
{\displaystyle{\left({\partial \over \partial \tau} \right)^n
    \left[ N_{l_0,l}(d) G_l/l \right]} \over
  \displaystyle{\left({\partial \over \partial \tau} \right)^n
    \left[ G_{l_0}/l_0 \right]} }\ .                         \label{DiscRho}
\end{equation}

\noindent
$N_{l_0,l}(d) G_l/l$ generates only surfaces with at least one boundary
loop of the desired sort.  Moreover, if there are $p$ such boundary
loops on a particular surface, $N_{l_0,l}(d) G_l/l$ will generate that
surface $p$ times: each time, the
tube generated by $N_{l_0,l}(d)$ will have a different one of its boundary
loops left open for $G_l/l$ to plug.  Hence $N_{l_0,l}(d) G_l/l$ is
the sum over surfaces generated by $G_{l_0}/l_0$ of the weight of
each surface times the number of boundary loops at distance $d$ and of
length $l$ on that surface.  To understand the presence of the puncture
operators $\partial / \partial \tau$, think for a moment of surfaces with
fixed area instead of fixed $\tau$.  To get to the fixed area representation,
we would separately carry out inverse Laplace transforms on the numerator and
denominator.  The large area behavior of the fixed area quantities would be
controlled by the leading singular term in the small $\tau$ expansion of the
fixed $\tau$ quantities.  Inserting enough puncture operators
$\partial / \partial \tau$ in the numerator and denominator of
\equno{DiscRho} to eliminate the leading analytic terms in $\tau$ thus
provides a convenient way to isolate the terms that survive in the
large area limit.
In order to obtain a generalization of \equno{DiscRho} to disks of finite
area, we would have to replace the numerator and denominator by their
inverse Laplace transforms.

The continuum limit of \equno{DiscRho} is obviously

\begin{equation}
\rho(L_0, L,D) =  \lim_{\tau\to 0}
    {\displaystyle{\left({\partial \over \partial \tau} \right)^n
      \left[ N(L_0,L,D) F(L)/L \right]} \over
    \displaystyle{\left({\partial \over \partial \tau} \right)^n
      \left[ F(L_0)/L_0 \right]} }\ .                       \label{ContRho}
\end{equation}

\noindent
Let us first consider the limit $L_0\to 0$ where the entrance loop is
shrunk to a point.

We expand

\begin{equation}
F_k(L, \tau)/L = {1 \over L^{k+3/2}} + O(\tau) + \ldots + O(\tau^{k-1}) +
   \gamma_2 L^{k - 5/2} \tau^{k-1/2} + O(\tau^k),        \label{FL}
\end{equation}

\noindent
where $\gamma_2$ is another numerical factor and we have adjusted
$F_k(L, \tau)/L$ by an overall multiplicative
factor for convenience.  In this expansion,
as in \equno{NLL}, only the $O(\tau^0)$ term and the leading nonanalytic
term are relevant to $\rho(L_0, L, D)$.  This is because both the numerator
$N(L_0,L,D) F(L)/L$ and the denominator $F(L_0)/L_0$ of \equno{ContRho}
have $k$ leading terms analytic in $\tau$, followed by an $O(\tau^{k-1/2})$
term.  The leading terms are deleted by making $k$ punctures.  If one
made fewer punctures, a positive overall power of $L_0$ would make
$\rho(L_0, L,D)$ identically~$0$.  If one made more punctures, the leading
nonanalytic term would still be dominant as $\tau \to 0$, so the final
result would be unchanged.

In the $\tau \to 0$, $L_0\to 0$ limit, the only finite physical quantities
in $\rho(L_0,L,D)$ are $L$ and $D$, so the only possible dimensionless
scaling parameter is $x = L / D^{1/\sigma}$.  Thus $\rho(L_0\to 0, L, D)$
is a function only of $x$, up to a dimensionful overall factor:

\begin{equation}
\rho(L_0 \to 0, L, D) = {1 \over D^{1/\sigma}}
   \left[ {\gamma_1 \over\gamma_2 \Gamma(\sigma)} x^{-\sigma-2}
      (2\sigma+1+ x) + {x^\sigma \over \Gamma(\sigma + 1)} \right] e^{-x}\ ,
                                                             \label{RhoSol}
\end{equation}

\noindent
where $\sigma=k-{3\over 2}$, and
the numbers $\gamma_1$ and $\gamma_2$ can be calculated explicitly:

\begin{equation}
\gamma_1 = {2^{3 - 2k} \over (6k - 5) (2k - 3)}\ ,\qquad\qquad \quad
\gamma_2 = {1 \over (2k-1)!! (2k-3)!!}\ .
\end{equation}

\noindent
The terms in square brackets in \equno{RhoSol} separate beautifully as
$k \to \infty$: the first term gives a non-integrable divergence at
$x = 0$, and the second term gives a Poisson distribution, normalized to
one and peaked at $x = \sigma$.

The first term shows that the random surface
has huge numbers of protuberances whose circumferential length is small
compared to their geodesic distance from a given point; this is what we
mean by the surface being hairy.  The profusion of ``microscopic''
boundary loops suggests that as a one-dimensional universe
propagates through geodesic time, it emits a divergent number of baby
universes.

For large $x$, the second term in square brackets is dominant, and it
shows that there is exactly one ``macroscopic'' boundary loop
to our region, and for large $k$ its length is sharply peaked about
$\sigma D^{1/\sigma}$.  Thinking again of a one-dimensional universe
propagating through geodesic time, we would interpret the one macroscopic
boundary loop as the parent universe, which survives the emission of
its numerous baby universes.

To determine at what $x$ the first term becomes significant, we note that

\begin{equation}
{1\over \sigma} \log \left[
   {\gamma_1\over \gamma_2\Gamma(\sigma)} x^{-\sigma-2}
      (2\sigma + 1 + x) e^{-x} \right]
  \to -(1+s+\log s)
\end{equation}

\noindent
as $\sigma = k-3/2 \to \infty$ with
$s \equiv x/\sigma$. This is positive for $s<s_c$ and negative for $s>s_c$,
where $s_c$ satisfies

\begin{equation}
  1+s_c+\log s_c=0\ ,
\end{equation}

\noindent
the solution to which is $s_c \approx 0.2784645428$.  Let us define
$\rho(s)$ by

\begin{equation}
\rho(s) \, ds = \rho(L_0 \to 0,L,D) \, dL =
  \left[ {\gamma_1\over \gamma_2\Gamma(\sigma)} x^{-\sigma-2}
    (2\sigma+1 + x) + {x^\sigma \over \Gamma(\sigma + 1)} \right] e^{-x}\, dx\
{}.
\end{equation}

\noindent
It follows from the above discussion that, as $k \to \infty$,
$\rho(s)$ converges in the weak sense
to a distribution which is $+\infty$ for $s < s_c$ and
$\delta (s - 1)$ for $s > s_c$.  In Fig.~6 we show plots of $\rho(s)$
for $k = 2$, where there is no separation of the ``microscopic'' and
``macroscopic'' terms; for $k=6$, where the separation is significant;
and for $k=100$, where the convergence to the limiting case is very clear.

Large $k$ corresponds to the central charge going to $-\infty$, which is
held to be the semiclassical limit where quantum fluctuations vanish and
the surface is smooth, with constant scalar curvature.  In the
thermodynamic (infinite area) limit, the surface would be a plane, and we
would have $\rho(L_0 \to 0,L,R) = \delta(L - 2\pi R)$, so
$\rho(s) = \delta(s - 1)$ with $s = L/(2 \pi R)$.  It is intriguing that
this classical term is present in semiclassical limit we found for
$\rho(s)$.  The different scaling law,

\begin{equation}
s = {L\over \sigma D^{1/\sigma}}\ ,                      \label{Slaw}
\end{equation}

\noindent
and the
profusion of microscopic boundary loops for $s < s_c$, are striking features
of the quantum case which we cannot conceive of predicting by quasi-classical
arguments.  The scaling law \equno{Slaw}
might seem to be an
artifact of the combinatorics: on the discrete surface, the matter fields
are incorporated into the manifold by polygons with different numbers of
sides, so the notion of defining geodesic distance as the minimal number of
polygons one must traverse to get from point to point is suspect.  So,
couldn't we just define
$R = \sigma D^{1/\sigma} / 2\pi$ and claim that $R$ is the ``real'' geodesic
distance?  The problem with this approach is that $D$ enjoys a linearity
property that seems essential to the notion of geodesic distance.  Namely,
if $\gamma$ is a loop each of whose points is a geodesic distance $D_1$ from
a given point $P$ on a random surface, and if $\gamma'$ is a loop each of
whose points is a geodesic distance $D_2$ from $\gamma$, then each point
on $\gamma'$ will be a geodesic distance $D_1 + D_2$ from $P$.  Any increasing
function of $D$ with the same property would have to be linear.  The
authors feel that a resolution of this question will have to come from a
continuum formalism where the matter fields are more easily distinguished
from the metric on the manifold.

Another interesting limit where exact calculations are possible is that
of an extremely long entrance loop, $L_0\to \infty$.
Here it is appropriate to calculate $N(\zeta,L',D)$ in the limit where
$\zeta^{-1}$ and $\tau^{-1/2}$ are much larger
than $D^{1/\sigma}$ and $L'$.  Now the characteristic curve equation,
\equno{IntEQ}, has a very simple approximate solution,

\begin{equation}
\zeta_0 \approx \zeta- Df_k(\zeta, \tau)\ ,               \label{SimpSol}
\end{equation}

\noindent
so that, from \equno{DSol},

\begin{equation}
N(\zeta,L',D) =1-\zeta L' + L'Df_k(\zeta, \tau)+\ldots    \label{Smallz}
\end{equation}

\noindent
Performing the inverse Laplace transform, we obtain

\begin{equation}
N(L\to\infty, L', D)= L' D F_k(L, \tau)\ .
\end{equation}

\noindent
Using this and \equno{FL} in \equno{ContRho}, we arrive at a
remarkably simple formula:

\begin{equation}
\rho(L_0\to\infty, L, D)= L_0 D L^{-k-1/2}\ .           \label{LargeL}
\end{equation}

\noindent
This distribution is valid whenever $\tau^{-1/2} \gg L_0$ and both are
much larger than $L$ and $D^{1/\sigma}$.  The fact that the number of emitted
\bu s scales as $L_0$ could have been anticipated: this is simply
due to the fact that a \bu\ can split off anywhere along the parent.
It is also clear why the number grows linearly with the elapsed
geodesic time $D$, since in the discrete case the number of \bu s that
split off at each step of evolution should be constant as long as the
length of the parent universe does not change appreciably.  Note that
the distribution in $L$ is again non-integrable for small $L$: the emitted
\bu s are overwhelmingly likely to be microscopic.

Another consequence of the preceding discussion is that

\begin{equation}
N(L, L'\to 0, D\to 0)= L' D F_k(L, \tau)\ .                \label{TubeD}
\end{equation}

\noindent
This formula is interesting because it establishes a connection between
the tube amplitude and the disk amplitude.  $N(L, L'\to 0, D\to 0)$ is the
sum over disks of boundary length $L$ with a marked point
located at a vanishing geodesic distance $D\to 0$ from the boundary.
Thus, we expect that in the $L' \to 0$, $D \to 0$ limit
the tube amplitude reduces to the disk amplitude with a marked
boundary point.  This is indeed what happens, according to \equno{TubeD}.
We speculate that \equno{TubeD} or some more general form of it could be used
as the basis of a continuum derivation of \equno{DiffEQ}.

\section{Discussion}
\label{Discuss}

In ref.~\cite{Kawai2} a string field theory formalism for $c=0$
gravity was introduced. {}From this formalism  \equno{DiffEQ} was
elegantly
derived. In fact, \equno{DiffEQ} was derived first in ref.~\cite{Kawai1}
via a careful combinatorial analysis of discretizations, and the
string field theory was tailored to reproduce this result. In this paper
we extended the combinatorial analysis to arbitrary discretizations
with $2m$-gons and established the validity of \equno{DiffEQ}
for 2-dimensional gravity coupled to the $(2, 2k-1)$ minimal models.
This strongly suggests that the string field theory formalism of
ref.~\cite{Kawai2} encompasses all these theories. We choose a
particular theory only through its disk amplitude
$f(\zeta, \tau)$, which is the background value of the string field.

In solving \equno{DiffEQ} we found a drastic difference between the even
and odd $k$. In other calculations no such major differences were noted.
For instance, the disk amplitudes of \equno{ExactFL} are positive for all $k$.
Thus, we may have the first indication of a serious difference between
even and odd $k$ occurring for spherical surfaces. It would be nice to
understand a deeper reason behind this.

The non-integrable divergence of $\rho(L,D)$ for small $L$ indicates that
random surfaces are very hairy, even when restricted to
spherical topology.  In terms of the propagation through geodesic distance
of a loop along its world-sheet, we take this to mean that tiny
loops---the baby universes---are constantly splitting off
the main loop. The conclusion that the microscopic \bu s are
overwhelmingly more likely to split off than the macroscopic ones
is quite intriguing in light of the large wormhole problem \cite{VK}.

One might think of the microscopic \bu s as analogous
to the soft photons which create the well-known infrared
problem in the bremsstrahlung cross-section.
Therefore, only the inclusive probabilities, where we sum over all
possible splittings, are non-vanishing.
We have verified that
the probability for a loop to propagate any finite $D$ along its world
sheet without splitting is zero.

The limit $k \to \infty$ (corresponding to the central charge decreasing
without bound) bears a subtle relationship to classical gravity
which we do not fully understand.  Suggestions of a manifold that is smooth
at length scales large compared to $D^{1/\sigma}$ emerge from the
$k \to \infty$ limit of $\rho(L_0 \to 0,L,D)$.  But baby universes still
play an important role in this limit, as evidenced by the divergence of
$\rho(L_0 \to 0,L,D)$ as $k \to \infty$ for $L < s_c \sigma D^{1/\sigma}$.
Perhaps the proliferation of \bu s up to this critical scale
is related to the presence of many
operators of negative dimension.

It would be very interesting to study models with unitary conformal
fields coupled to gravity, to see whether the surfaces they produce are
more regular or more wild.  To calculate the transfer
matrix with the same methods as described here, however, one would need
the disk amplitudes of the model with arbitrary boundary conditions on
the matter fields, and these are not available even for such simple models
as the Ising model.

\section*{Acknowledgements}

We thank A. Polyakov and N. Seiberg for useful discussions.
This work was supported in part by DOE grant DE-AC02-76WRO3072,
the NSF Presidential Young Investigator Award PHY-9157482,
James S. McDonnell Foundation grant No. 91-48,
and an A. P. Sloan Foundation Research Fellowship.

\section*{Figures}

\noindent
1. Small example of a one-step evolution of a loop on a random
surface tiled with squares.

\noindent
2. Basic shapes for $k=2$.

\noindent
3. The first few terms contributing to one basic shape, drawn
on both the polygonal and dual lattices.

\noindent
4. The composition law.

\noindent
5. Characteristic curves for $k=3$ (top left) and $k=4$ (top right)
above graphs of $f_3(\zeta,\tau)$ (bottom left) and $f_4(\zeta,\tau)$
(bottom right).

\noindent
6. The scaling function $\rho(s)$ for $k=2$, $k=6$, and $k=100$.

\end{document}